# Strong Rashba-Edelstein Effect-Induced Spin-Orbit Torques in Monolayer Transition Metal Dichalcogenides/Ferromagnet Bilayers

Qiming Shao[1*□], Guoqiang Yu[1*□], Yann-Wen Lan[1,2*], Yumeng Shi[3,5], Ming-Yang Li[3,4], Cheng Zheng[1], Xiaodan Zhu[1], Lain-Jong Li[3], Pedram Khalili Amiri[1] and Kang L. Wang[1□]

[1]*Device Research Laboratory, Department of Electrical Engineering, University of California, Los Angeles, 90095, USA*

[2]*National Nano Device Laboratories, Hsinchu 30078, Taiwan*

[3]*Physical Science and Engineering Division, King Abdullah University of Science and Technology (KAUST), Thuwal, 23955-6900, Kingdom of Saudi Arabia*

[4]*Research Center for Applied Sciences, Academia Sinica, Taipei 10617, Taiwan*

[5]*SZU-NUS Collaborative Innovation Center for Optoelectronic Science & Technology and Key Laboratory of Optoelectronic Devices and Systems of Ministry of Education, College of Optoelectronic Engineering, Shenzhen University, Shenzhen 518060, China*

*These authors contributed to this work equally.

□corresponding authors: sqm@ucla.edu, guoqiangyu@ucla.edu, wang@ee.ucla.edu

The electronic and optoelectronic properties of two-dimensional materials have been extensively explored in graphene and layered transition metal dichalcogenides (TMDs). Spintronics in these two-dimensional materials could provide novel opportunities for future electronics, for example, efficient generation of spin current, which should enable the efficient manipulation of magnetic elements. So far, the quantitative determination of charge current-induced spin current and spin-orbit torques (SOTs) on the magnetic layer adjacent to two-dimensional materials is still lacking. Here, we report a large SOT generated by current-induced spin accumulation through the Rashba-Edelstein effect in the composites of monolayer TMD ($MoS_2$ or $WSe_2$)/CoFeB bilayer. The effective spin conductivity corresponding to the SOT turns out to be almost temperature-independent. Our results suggest that the charge-spin conversion in the chemical vapor deposition-grown large-scale monolayer TMDs could potentially lead to high energy efficiency for magnetization reversal and convenient device integration for future spintronics based on two-dimensional materials.

**KEYWORDS:** Spin-orbit torque, transition metal dichalcogenides, Rashba-Edelstein effect, spintronics, charge-spin conversion, two-dimensional materials

The efficient generation of spin current is crucial for improving the energy efficiency of spintronics. The spin current can be used to exert spin-orbit torques (SOTs) on a magnetic layer, enabling the manipulation and even switching of magnetization in an energy efficient way[1, 2]. In the past decade, heavy metals, such as Pt[3, 4] and Ta[5-8], or bulk semiconductors, such as GaAs[9, 10] have been extensively studied due to the presence of a strong spin-orbit coupling, allowing the spin Hall effect or the Rashba-Edelstein effect (REE) to generate a spin accumulation. Compared with these three-dimensional materials, the conversion between spin and charge in two-dimensional materials, *i.e.*, van der Waals materials, has not been studied until recently[11].

Monolayer graphene has been extensively studied as a spin channel due to its weak spin-orbit coupling[12]. A modified graphene with an enhanced spin-orbit coupling strength or increased extrinsic spin-dependent scattering rates could give rise to a significant spin Hall effect. However, it requires specific treatments, such as hydrogen bonding[13] or Cu (Au) adatoms[14], which are hard to control. A giant SOT was demonstrated in heterostructures based on three-dimensional topological insulators, *i.e.*, the $Bi_2Se_3$ family, which are also van der Waals materials[15-17]. The colossal SOT originates from the topological surface states. However, a thickness larger than the hybridization length of two surface states, six quintuple layers (~ 6 nm), is needed for topological insulators[18]. So far, it remains elusive whether or not we can have a large spin torque from an ultrathin atomically monolayer film (< 1 nm). Monolayer transition metal dichalcogenides (TMDs), such as $MX_2$ (M = Mo, W, X = S, Se, Te), provide a unique platform for studying the generation of SOTs at the two-dimensional limit because monolayer TMDs have both strong spin-orbit coupling and inversion symmetry breaking[19-22]. Very recently, signatures of current-induced SOTs were found in the composite of monolayer $MoS_2$/ferromagnet bilayer[23], but the SOTs have not been quantitatively characterized and the origin of the SOTs has not been interpreted.

In this Letter, we report the observation of current-induced SOTs in $MX_2$/CoFeB bilayers, where the $MX_2$ is monolayer $MoS_2$ or $WSe_2$. The monolayer $MX_2$ is grown by chemical vapor deposition (CVD) and has a size up to mm scale. Using a second-harmonic method, we succeeded in determining both field-like torque per unit moment (or in-plane spin-orbit effective field) and damping-like torque per unit moment (or out-of-plane spin-orbit effective field). The field-like torque is large in $MX_2$/CoFeB bilayers despite most of the current are going through the CoFeB layer. The damping-like torque is negligible within measurement uncertainty, which is consistent with the REE dominated SOT generation in the $MX_2$/CoFeB bilayers. Moreover, the current-induced in-plane spin conductivity due to the REE is almost independent of temperature.

High-quality large-area monolayer $MoS_2$ and $WSe_2$ were grown on sapphire using CVD method, where the transition metal trioxides were vaporized and reacted with the S or Se vapor in a chamber under a controlled temperature and gas environment (see details in Supporting Information Section S1)[24, 25]. The insets of Figs. 1a and 1b are the optical images of $MoS_2$ and $WSe_2$, which show the uniformity of the thin film sample. Raman spectra further confirm that the films are monolayers. The Raman spectrum of $MoS_2$ (see Fig. 1a) exhibits two characteristic bands: the in-plane phonon mode, $E_{2g}^1$, centered near 385 cm$^{-1}$ and the out-of-plane phonon mode, $A_{1g}$, centered near 405 cm$^{-1}$, with a peak frequency difference of 20 cm$^{-1}$, which is a clear signature of monolayer $MoS_2$. Similarly, a high-intensity peak ($E_{2g}^1$) shows near 250cm$^{-1}$ for $WSe_2$ (see Fig. 1b), which indicates that the $WSe_2$ film is a monolayer as well. The sheet resistances of monolayer $MoS_2$ and $WSe_2$ are larger than $10^6$ $\Omega$/sq as shown in the current-voltage curve (see Fig. 1c). To study the current-induced SOTs on the magnetic moment, we deposited 3 nm CoFeB on top of the monolayer $MoS_2$ and $WSe_2$ using a magnetron sputtering system. The deposition rates were 0.03 nm/s for CoFeB in an argon pressure of 3 mTorr. The CoFeB layer was capped by $TaO_x$ (~ 3 nm). For details of the deposition process and Raman characterization of $MX_2$/CoFeB bilayers after the deposition, see Supporting Information

Section S2. The MX$_2$/CoFeB bilayers were patterned into Hall bars (channel width is 20 μm) using standard photolithography as shown in Fig. 1d. We used a second-harmonic analysis of both anomalous Hall resistance and planar Hall resistance ($R_{\text{Hall}}^{2\omega} = V_{\text{Hall}}^{2\omega}/I_{\text{ac\_peak}} = R_{\text{AHE}}^{2\omega} + R_{\text{PHE}}^{2\omega}$) to determine the current-induced spin-orbit effective fields in the MX$_2$/CoFeB bilayers as in refs. [26, 27]. The applied a.c. current frequency is $\frac{\omega}{2\pi} = 35.85$ Hz. Since the magnitudes of second-harmonic signals are proportional to the a.c. current amplitude, here we only present the results using an a.c. current amplitude 1 mA (*r.m.s.* value).

We examine the magneto-transport properties of the MX$_2$/CoFeB bilayers using a physical properties measurement system at T = 300 K unless otherwise stated. In the following, we will first present the results for the MoS$_2$/CoFeB bilayer, and subsequently the WSe$_2$/CoFeB bilayer when we discuss the results. The MoS$_2$/CoFeB bilayer shows an in-plane easy plane, and the effective anisotropy field ($H_K$) is -1 T (see Fig. 2a). Here, we define the perpendicular magnetic anisotropy by a positive value of the anisotropy field. There is no easy axis in the film plane because the planar Hall resistance as a function of in-plane azimuthal angle ($\varphi$) in the presence of in-plane external magnetic field ($H_{\text{ext}} = 1000$ Oe) follows the sin $2\varphi$ relation well as shown in Fig. 2b. The presence of monolayer MoS$_2$ layer does not affect the magnetic properties of the 3nm CoFeB layer since the MoS$_2$/CoFeB bilayer has similar saturation anomalous Hall resistance ($R_A$), planar Hall resistance ($R_P$), and effective anisotropy field as the 3 nm CoFeB directly deposited on the SiO$_2$ (see the Supporting Information Table S1).

The idea of SOT measurement is described as following. When the injection charge current passes through the MoS$_2$/CoFeB bilayer, a net spin accumulation could develop in a direction transverse to the current direction in the film plane due to the inversion symmetry breaking and spin-orbit coupling in the monolayer MoS$_2$ and/or the MoS$_2$/CoFeB bilayer. In other words, the nonequilibrium spin accumulation $\boldsymbol{\sigma} \propto \boldsymbol{z} \times \boldsymbol{j}$, where the mirror symmetry with respect to the *xy* plane is broken, and $\boldsymbol{j}$ is the current direction (along $\pm y$ direction). This current-induced spin polarization, in general, could give rise to two types of SOTs, the field-like torque ($\boldsymbol{\tau_F} = \boldsymbol{m} \times \boldsymbol{\sigma}$) and the damping-like torque ($\boldsymbol{\tau_D} = \boldsymbol{m} \times (\boldsymbol{m} \times \boldsymbol{\sigma})$). Therefore, the in-plane azimuthal angle dependence of $R_{\text{Hall}}^{2\omega}$ can be divided into two major components depending on the symmetry of current-induced SOTs (see Fig. 2c and the Supporting Information Section S3):

$$R_{\text{Hall}}^{2\omega} = R_\parallel \cos 2\varphi \sin \varphi + R_\perp \sin \varphi = R_{\text{P}} \frac{H_\parallel}{|H_{\text{ext}}|} \cos 2\varphi \sin \varphi + \frac{R_{\text{A}}}{2} \frac{H_\perp}{|H_{\text{ext}}| - H_{\text{K}}} \sin \varphi, \qquad (1)$$

where the first term originates from the current-induced in-plane spin-orbit effective field ($\boldsymbol{H_\parallel}$) and the second term comes from the current-induced out-of-plane spin-orbit effective field ($\boldsymbol{H_\perp}$). The magnitudes of both in-plane and out-of-plane spin-orbit fields are proportional to the magnitude of current. When the current is along the +*y* axis, the $\boldsymbol{H_\parallel}$ is along the -*x* axis and is independent with the magnetization. Therefore, the $\boldsymbol{H_\parallel}$ gives rise to a field-like torque $\boldsymbol{\tau_F} = \boldsymbol{m} \times \boldsymbol{H_\parallel}$. This oscillating $\boldsymbol{H_\parallel}$ induced by the a.c. current causes the magnetization to oscillate in the film plane, and thus gives rise to a $R_{\text{PHE}}^{2\omega} \propto \cos 2\varphi \sin \varphi$ since $\frac{dR_{\text{PHE}}}{d\varphi} \propto \cos 2\varphi$ and $H_\parallel \propto \sin \varphi$. As shown in Fig. 2c, $R_{\text{PHE}}^{2\omega}$ reaches minimum $-R_\parallel$ when the $\boldsymbol{H_{\text{ext}}}$ is along the +*y* direction, and maximum $R_\parallel$ when the $\boldsymbol{H_{\text{ext}}}$ is along the -*y* direction. The field dependence of $R_\parallel$ follows $1/|H_{\text{ext}}|$, which is consistent with the picture of an in-plane spin-orbit field; the larger the external field is, the smaller angle the current-induced field-like torque can induce. By fitting the field dependence of the extracted $R_\parallel$ with $R_{\text{P}} \frac{H_\parallel}{|H_y|}$, where $H_y$ is the magnetic field along the $\pm y$ direction, we obtain the magnitude of $H_{\parallel,\text{MoS}_2}/I_{\text{ac\_peak}} \approx 0.13$ Oe/mA (see Fig. 2d). To get a more intrinsic property for the SOT generation, we convert the $H_\parallel$ into the effective spin conductivity ($\sigma_\parallel$) using $\sigma_\parallel = \frac{J_s}{\mathcal{E}} = \frac{H_\parallel M_s t_{\text{FM}}}{\mathcal{E}}$ [16], where $M_s t_{\text{FM}}$ is the saturation magnetization per unit area for the 3 nm CoFeB layer, and $\mathcal{E}$ is the applied electric field inside the MoS$_2$/CoFeB bilayer. We independently determine $M_s t_{\text{FM}} = 2.34$ mA using superconducting quantum interference device and $\mathcal{E} = 3.21 \times 10^4$ V/m for $I_{\text{ac\_peak}} = 1$ mA. Therefore, the corresponding effective in-plane spin conductivities for MoS$_2$/CoFeB is $\sigma_{\parallel,\text{MoS}_2} \approx 2.88 \times 10^3 \hbar/2e$ ($\Omega^{-1}\text{m}^{-1}$). If we consider that the electronic conductivity of the monolayer MoS$_2$ is very low, the intrinsic ratio of generated spin current density over charge current density, or the so-called "spin torque ratio," will be relatively high, and comparable with traditional heavy metals, such Pt and Ta (will be discussed below).

The current-induced $\boldsymbol{H}_\perp$ changes its direction when the direction of magnetization is reversed, and thus it gives rise to a damping-like torque $\boldsymbol{\tau_D} = \boldsymbol{m} \times \boldsymbol{H}_\perp$. The $\boldsymbol{H}_\perp$ induced by the a.c. current causes the magnetization to oscillate out of the film plane and thus gives rise to a $R_{\text{AHE}}^{2\omega} \propto \sin\varphi$ since $\frac{dR_{\text{AHE}}}{d\theta}|_{\theta=90°} = R_A \frac{d\cos\theta}{d\theta}|_{\theta=90°} = R_A$ and $H_\perp \propto \boldsymbol{m} \times \boldsymbol{\sigma} \propto \sin\varphi$. The $R_\perp$ decreases as the external magnetic field increases according to Eq. (1). As shown in Fig. 3a, we do see a distinct field dependence of the extracted $R_\perp$ for the Ta(0.8 nm)/CoFeB bilayer (as the control sample), and the estimated $H_{\perp,\text{Ta}}/J_{\text{ac\_peak}}$ is around 2.71 Oe per $10^{11}$ A/m$^2$. This value is consistent with the previously reported values on the ultrathin Ta film with the same thickness[28]. In addition to the field-dependent $R_{\text{AHE}}^{2\omega}$, there is an additional step function $R_{\text{Hall},\nabla\text{T}}^{2\omega}$ as illustrated in the inset of Fig. 3a; the $R_{\text{Hall},\nabla\text{T}}^{2\omega}$ changes sign as the magnetization direction reverses, while the magnitude does not vary with that of the external magnetic field. This step function is due to the anomalous Nernst effect (ANE), a thermoelectric effect[27] (see the Supporting Information Section S4). Nevertheless, we can differentiate the thermo-voltage ($V_{\text{Hall},\nabla\text{T}}^{2\omega}$) and the SOT-induced second-harmonic anomalous Hall resistance ($R_{\text{Hall}}^{2\omega}$) by their field dependencies. For the MoS$_2$/CoFeB bilayer, as shown in Fig. 3b, within measurement uncertainty, we do not observe a clear trend that the $R_\perp$ decreases as field increases (see Supporting Information Section S5 for more information about the measurement uncertainty due to the choice of fitting range of the magnetic field). The observed negative $V_{\text{Hall}}^{2\omega}$ when the magnetization is along the +$y$ direction is consistent with the ANE picture (see Fig. 3b). When the magnetization reverses, the $V_{\text{Hall}}^{2\omega}$ becomes positive as expected from the ANE. So, the ANE dominates in the observed $V_{\text{Hall}}^{2\omega}$, and the damping-like torque or $H_{\perp}$ is not observed within measurement uncertainty.

Here, we interpret that the REE is the mechanism for the observations of a large $H_{\parallel}$ and a negligible $H_{\perp}$ in the sapphire/MoS$_2$/CoFeB/TaO$_x$ heterostructure (see the Supporting Information Section S6). The REE appears in the presence of a electric potential gradient and strong spin-orbit coupling, *i.e.*, typically at the interface between a material with strong spin-orbit coupling and a different material, such as the Bi/Ag interface[29]. In our MoS$_2$/CoFeB bilayer, the strong spin-orbit coupling and the broken vertical symmetry (together with the intrinsic inversion symmetry breaking in the monolayer MoS$_2$) could give rise to a large Rashba-type spin splitting[30]. The Rashba Hamiltonian can be expressed by $H_R = \alpha_R(\boldsymbol{k}\times\boldsymbol{z})\cdot\boldsymbol{\sigma}$, where $\alpha_R$ is the Rashba coefficient, $\boldsymbol{k}$ is the electron momentum, and $\boldsymbol{\sigma}$ is the spin Pauli matrices. As shown in Fig. 1e, at the equilibrium state, there is no net spin accumulation due to an equal number of electrons moving in two directions. Under an external electric field along the +*y* direction, the Rashba spin-split Fermi surfaces shift, causing a net spin accumulation along the -*x* direction, which is consistent with the direction of the observed in-plane spin-orbit field. Moreover, theoretical calculation shows that to the first order, the Rashba spin-splitting can only give rise to a field-like torque[31] or the Rashba effect gives a much larger field-like torque compared with the damping-like torque[32] as we observed in the MoS$_2$/CoFeB bilayer. If the spin Hall effect plays an important role in the MoS$_2$/CoFeB, we should have seen a sizeable damping-like torque like the Ta/CoFeB (see Table 1). However, we didn't observe any significant damping-like torque in the MoS$_2$/CoFeB. Regarding the charge-spin conversion efficiency, it has been shown that the inverse REE can convert the spin current into the charge current, and the efficiency is quantified as $\lambda_{\text{IREE}} = \alpha_R\tau_s/\hbar$ [29], where $\tau_s$ is the effective spin relaxation time and $\hbar$ is the reduced Planck constant. Since the valley and spin are coupled in monolayer MoS$_2$, the relaxation time of spin polarization could be longer than 1 ns due to the considerable energy required for flipping the valley index[33]. So, the charge-spin conversion efficiency could be very high in the MoS$_2$/CoFeB bilayer.

From the harmonic measurement, we learn that the in-plane effective spin conductivity is $\sigma_{\parallel,\text{MoS}_2} \approx 2.88 \times 10^3 \hbar/2e$ ($\Omega^{-1}\text{m}^{-1}$), even when most of the current does not flow through the MoS$_2$ layer. The conductivity of MoS$_2$ in the MoS$_2$/CoFeB bilayer is around $\sigma_{\text{MoS}_2} \approx 2.1 \times 10^4$ ($\Omega^{-1}\text{m}^{-1}$) (assuming that the thickness of monolayer MoS$_2$ with van der Waals gaps is 0.8 nm) and thus the spin torque ratio, *i.e.*, the ratio of spin current density over charge current density, is given by $\vartheta_{\parallel,\text{MoS}_2} = \frac{2e}{\hbar} \sigma_{\parallel,\text{MoS}_2} / \sigma_{\text{MoS}_2} \approx 0.14$. If we can find an intrinsic MoS$_2$ with a much higher resistivity ($>10^6$ $\Omega$/sq), for example, by putting a monolayer MoS$_2$ on top of a magnetic insulator, and assume that the $\sigma_{\parallel,\text{MoS}_2}$ remains the same, an even larger spin torque ratio ($>2.3$) could be obtained. A recent experiment on the spin-charge conversion, the Onsager reciprocal process of the charge-spin conversion, in the Co/Al/MoS$_2$ heterostructure shows that the efficiency of the spin-charge conversion is very high, and the estimated $\lambda_{\text{IREE}}$ can be as large as 4.3 nm, which corresponds to a spin torque ratio as large as 12.7 [34]. More recently, spin-torque ferromagnetic resonance in the MoS$_2$/Permalloy bilayer reveals a large symmetric Lorentzian peak compared with the antisymmetric Lorentzian peak[23], which could be ascribed to either a large damping-like torque or a highly efficient spin pumping-driven inverse REE. Combining the results given by the spin-torque ferromagnetic resonance measurement[23] and the present work, we can claim that the large symmetric Lorentzian peak is mainly due to the inverse REE induced by the spin pumping, rather than the damping-like torque generated by the rf current.

To see if other TMDs can produce such a large in-plane spin-orbit field, we carry out the SOT measurement on another TMD material, WSe$_2$. The extracted $R_{\parallel}$ and $R_{\perp}$ as a function of an external magnetic field along $\pm y$ direction are plotted in Fig. 4a and Fig. 4b, respectively. Similar to the $MoS_2$/CoFeB bilayer, the in-plane spin-orbit field $H_{\parallel,WSe_2}/I_{ac\_peak} \approx 0.19$ Oe/mA, and we do not observe the damping-like out-of-plane spin-orbit field within measurement uncertainty. Using the same conversion method, we determine the effective in-plane spin conductivity $\sigma_{\parallel,WSe_2} \approx 5.52\times10^3 \hbar/2e$ ($\Omega^{-1}m^{-1}$), which is larger than the $\sigma_{\parallel,MoS_2}$ and is consistent with the stronger spin-orbit coupling in the monolayer $WSe_2$ compared with the $MoS_2$[21]. However, we should notice that although the monolayer $MoS_2$ and $WSe_2$ have very different conductivity (the monolayer $MoS_2$ has much higher resistivity than the monolayer $WSe_2$ in our study), they have similar spin conductivity. This result indicates that spin torques in these bilayers share the same origin, *i.e.*, REE[35].

We also study the temperature dependence of the current-induced in-plane spin conductivity. We do not identify the damping-like torque within the investigated temperature range. We observed that the current-induced in-plane spin conductivity is almost temperature independent (slightly increases as the temperature decreases) as shown in Fig. 5, which is similar to the report on the inverse REE in the Ag/Bi interface[36]. A possible explanation for the temperature-independent charge-spin conversion due to REE is the temperature-insensitive strength of Rashba spin-splitting and the Fermi level position. The Rashba spin-splitting developed at the $MX_2$/CoFeB interface relies on the band structure or wave function hybridization between the $MX_2$ and CoFeB[30, 36]. The band structure and the Fermi level position of $MX_2$/CoFeB could be temperature independent as reflected in the temperature independence of resistance (slight increase as the temperature decreases) of $MX_2$/CoFeB bilayers (see the inset of Fig. 5). However, detailed theories and more experiments are still required to fully understand the results presented in this paper.

In conclusion, we have shown that a current can generate a large in-plane spin-orbit effective field in a bilayer consisting of CVD-grown large-scale monolayer TMDs and a ferromagnetic layer and this effective field is temperature-insensitive. Our findings could be beneficial for future design of spintronic devices exclusively based on two-dimensional materials, where monolayer TMDs are coupled with magnetic van der Waals materials to form heterostructures that provide novel functionalities beyond electronics and optoelectronics[11, 37]. For future studies of two-dimensional semiconducting TMDs, on the one hand, systematic measurements on various TMDs need to be carried out to clarify the relation between the spin-orbit coupling strength and the spin torque efficiency. On the other hand, if the ferromagnetic metal we used in this study can be replaced by a magnetic insulator, such as yttrium iron garnet, there will be no shunting problem and the spin-charge conversion efficiency may be significantly improved[16]. Alternatively, metallic TMDs, such as 1T' phase WTe$_2$, have also been shown to give rise to a unique out-of-plane damping-like torque due to breaking of the in-plane mirror symmetry[38], which is preserved in our 1H phase TMDs. At last, we would like to mention that the REE-induced spin polarization at the atomically thin interface is expected to have a broad tunability with an external gate voltage[39], thus allowing for further improvement of energy efficiency for spintronic devices based on two-dimensional materials.

**Acknowledgements**

We thank Haojun Zhang, Dan Wilkinson and Bruce Dunn for discussions and assistance with experiments. Also, we thank the four anonymous reviewers whose comments and suggestions helped improve and clarify this manuscript. This work is supported as part of the Spins and Heat in Nanoscale Electronic Systems (SHINES), an Energy Frontier Research Center funded by the US Department of Energy (DOE), Office of Science, Basic Energy Sciences (BES), under Award # DE-SC0012670. We are also very grateful to the support from the Function Accelerated nanoMaterial Engineering (FAME) Center and Center for Spintronic Materials, Interfaces and Novel Architectures (C-SPIN), two of six centers of Semiconductor Technology Advanced Research network (STARnet), a Semiconductor Research Corporation (SRC) program sponsored by Microelectronics Advanced Research Corporation (MARCO) and Defense Advanced Research Projects

Agency (DARPA). Lain-Jong Li acknowledges the support from King Abdullah University of Science and Technology (Saudi Arabia), Ministry of Science and Technology (MOST) and Taiwan Consortium of Emergent Crystalline Materials (TCECM).

**Supporting Information Available**

The Supporting Information is available free of charge via the Internet at http://pubs.acs.org.

Details of the chemical vapor deposition of $MoS_2$ and $WSe_2$, deposition details and Raman characterization after deposition, principles of the second-harmonic measurement, effect of anomalous Nernst effect on the second-harmonic measurement, discussions on the determination of out-of-plane spin-orbit field, discussions on the origin of the current-induced in-plane spin-orbit field, properties of the investigated films in this work.

**Note:** The authors declare no competing financial interest.

**Figures and captions**

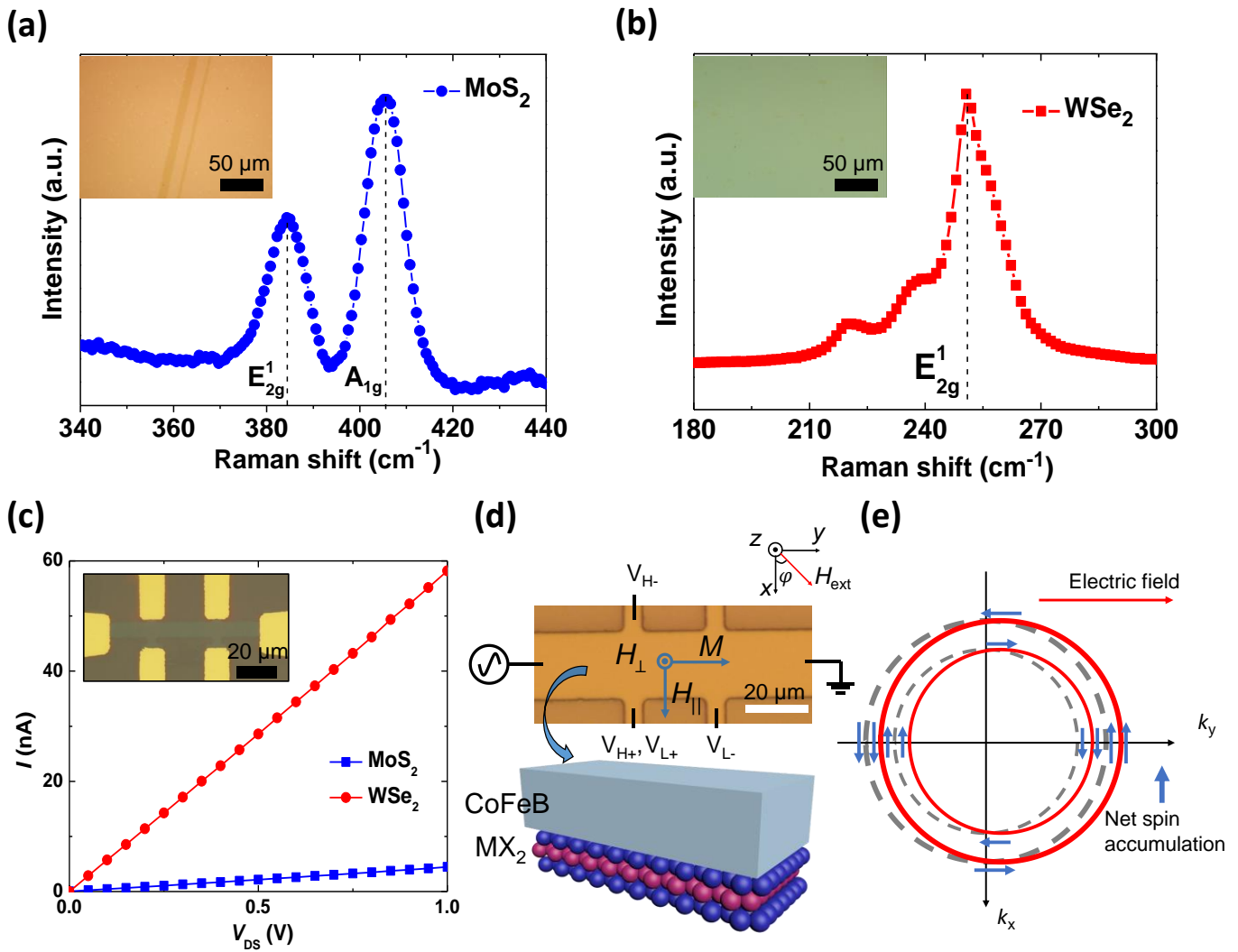

**Figure 1.** Materials characterization and measurement setup. Raman spectra of monolayer MoS$_2$ (a) and WSe$_2$ (b). The inset is a large scale optical image of monolayer MoS$_2$ (a) and WSe$_2$ (b) on sapphire. The scratches in (a) reveal the color contrast between the monolayer MoS$_2$ and the substrate. (c) Current-voltage characteristics of monolayer MoS$_2$ and WSe$_2$. The inset is an optical image of Hall bar structure used for the measurement. (d) Measurement setup of spin-orbit torque measurements for the MX$_2$/CoFeB bilayer. The MX$_2$ is a single layer, and the thickness of the CoFeB layer is 3 nm. (e) Illustration of induced spin accumulation by the Rashba-Edelstein effect at the interface of MX$_2$/CoFeB under an external electric field.

The dashed gray circles are Rashba spin-split Fermi surfaces in the equilibrium, and the solid red circles are for under an applied electric field. The blue arrows represent the spin angular momenta.

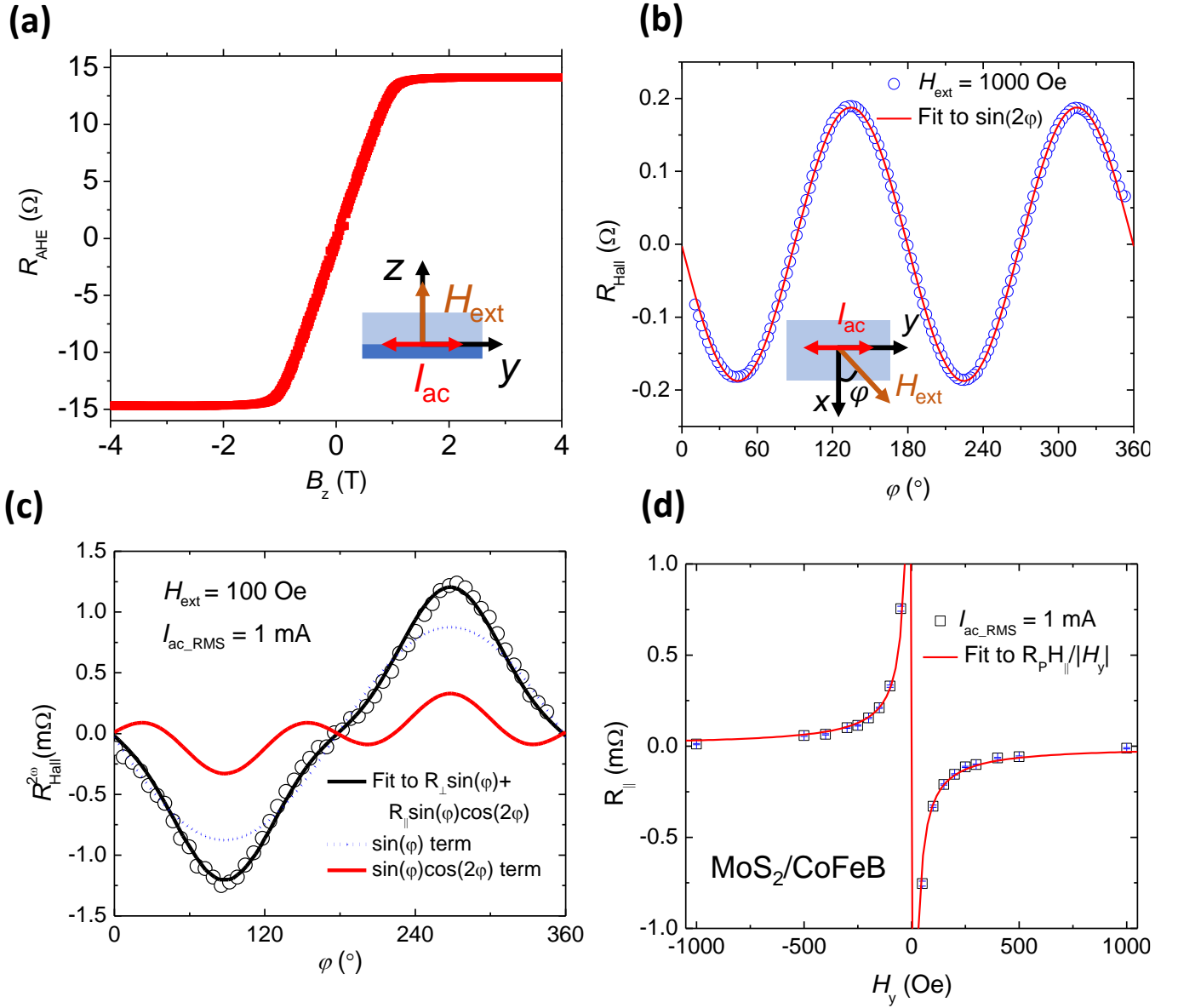

**Figure 2.** Field-like torque in the MoS$_2$/CoFeB bilayer. (a) Anomalous Hall resistance as a function of out-of-plane magnetic field. (b) Hall resistance as a function of in-plane azimuthal angle ($\varphi$) with an external magnetic field 1000 Oe applied. (c) Second-harmonic Hall resistance as a function of $\varphi$ with an external magnetic field 100 Oe applied. The black solid curve is fitted curve using $R_\perp \sin\varphi + R_\parallel \cos 2\varphi \sin\varphi$, where the first and second term are plotted in blue dotted and red solid curves, respectively. (d) The extracted $R_\parallel$ as a function of the external magnetic field along the $\pm y$ direction. The red solid curve is fitted curve using $R_P \frac{H_\parallel}{|H_y|}$, where the (field-like) in-plane spin-orbit field $H_\parallel$ is determined to be 0.18 Oe.

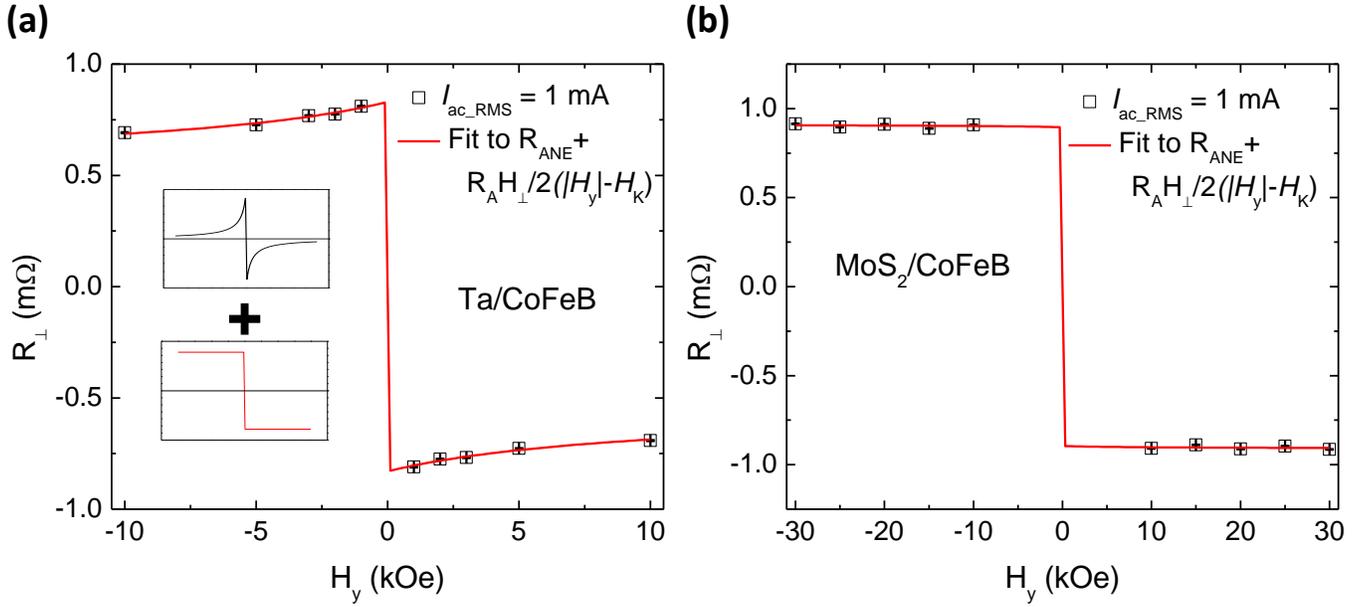

**Figure 3.** Determination of damping-like torque in the MoS$_2$/CoFeB bilayer. The extracted $R_\perp$ as a function of an external magnetic field along the $\pm y$ direction for Ta/CoFeB (a) and MoS$_2$/CoFeB (b). The red solid curves are fitted curves using $\frac{R_A}{2}\frac{H_\perp}{|H_y|-H_K} + R_{ANE}$, where the (damping-like) out-of-plane spin-orbit fields $H_\perp$ are determined to 0.50 Oe and negligible for the Ta/CoFeB and the MoS$_2$/CoFeB, respectively. In the inset of (a), field dependencies of damping-like torque term ($\frac{R_A}{2}\frac{H_\perp}{|H_y|-H_K}$) and anomalous Nernst Effect term ($R_{ANE}$) term are plotted on the top and bottom, respectively.

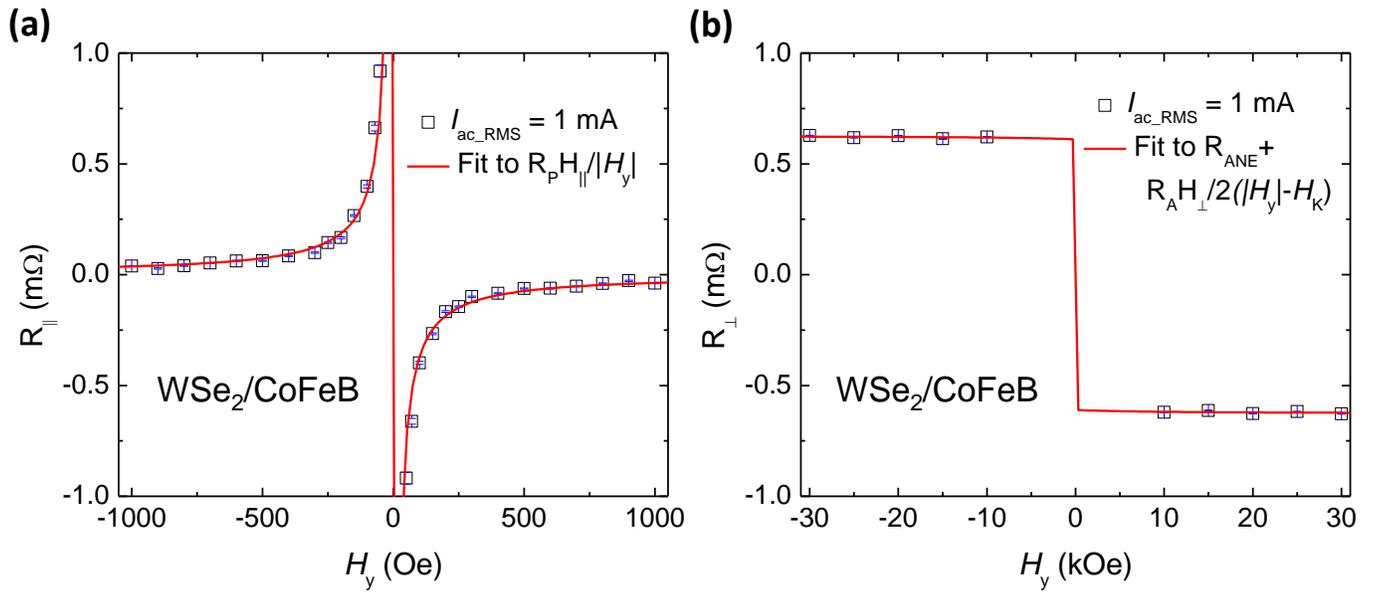

**Figure 4.** The extracted $R_{\parallel}$ (a) and $R_{\perp}$ (b) as a function of an external magnetic field along the $\pm y$ direction for the WSe$_2$/CoFeB bilayer.

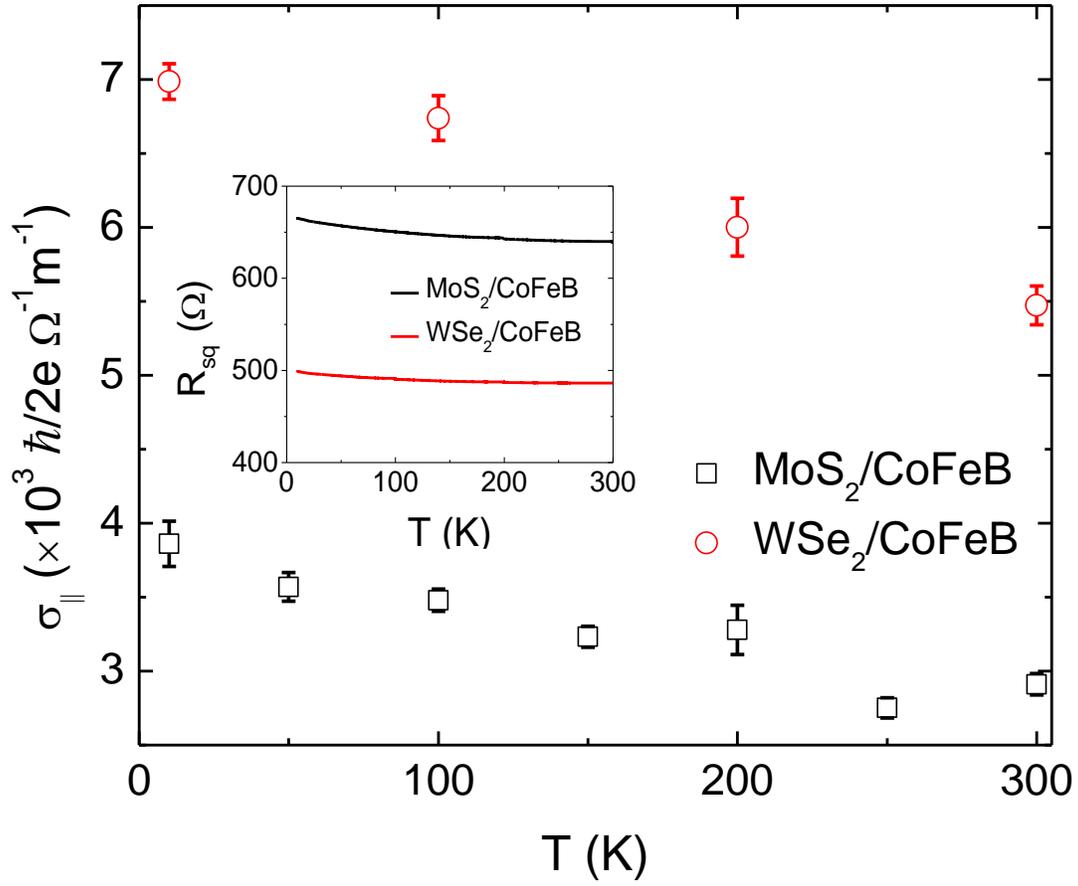

**Figure 5.** Temperature dependence of current-induced in-plane spin conductivities for both MoS₂/CoFeB and WSe₂/CoFeB. The inset shows the temperature dependence of sheet resistance for both MoS₂/CoFeB and WSe₂/CoFeB.

**Table 1. Current-induced spin-orbit fields in all the devices**

| Devices (nm) | MoS$_2$/CoFeB (3) | WSe$_2$/CoFeB (3) | Ta (0.8)/CoFeB (3) |
|---|---|---|---|
| In-plane (field-like) spin-orbit field ($H_\parallel$, Oe/mA) | 0.13 | 0.19 | 0.15 |
| Out-of-plane (damping-like) spin-orbit field ($H_\perp$, Oe/mA) | ~ 0 | ~ 0 | 0.35 |

**TOC graphic:**

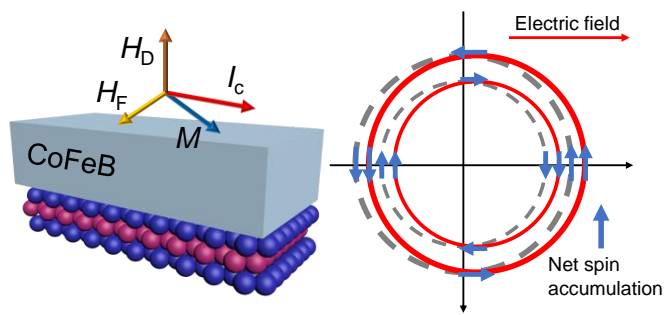